\documentclass[aps,prl,twocolumn,epsf,epsfig,amsmath]{revtex4}
\usepackage{CJK}
\usepackage[toc,page]{appendix}
\usepackage{hyperref}
\usepackage{graphicx,pdflscape,placeins}
\usepackage{color}
\usepackage{bm}
\usepackage{alltt,dsfont}
\usepackage{appendix}
\usepackage{amsmath,amssymb}
\usepackage{braket}
\usepackage[capitalise]{cleveref}
\usepackage{ulem}
\usepackage{float}

\begin{document}

\title{Orbital structure of the effective pairing interaction in the high-temperature superconducting cuprates}

\author{Peizhi Mai$^1$, Giovanni Balduzzi$^2$, Steven Johnston$^{3,4}$, Thomas A. Maier$^{1,5,*}$}
\affiliation{$^1$Center for Nanophase Materials Sciences, Oak Ridge National Laboratory, Oak Ridge, TN 37831-6164, USA}
\affiliation{$^2$Institute for Theoretical Physics, ETH Z{\"u}rich, 8093 Z{\"u}rich, Switzerland}
\affiliation{$^3$Department of Physics and Astronomy, University of Tennessee, Knoxville, Tennessee 37996-1200, USA}
\affiliation{$^4$Joint Institute for Advanced Materials at The University of Tennessee, Knoxville, TN 37996, USA}
\affiliation{$^5$Computational Sciences and Engineering Division, Oak Ridge National Laboratory, Oak Ridge, TN, 37831-6494, USA}
\date{\today}

\begin{abstract}
The nature of the effective interaction responsible for pairing in the high-temperature superconducting cuprates remains unsettled. This question has been studied extensively using the simplified single-band Hubbard model, which does not explicitly consider the orbital degrees of freedom of the relevant CuO$_2$ planes. Here, we use a dynamic cluster quantum Monte Carlo approximation to study the orbital structure of the pairing interaction in the three-band Hubbard model, which treats the orbital degrees of freedom explicitly. We find that the interaction predominately acts between neighboring copper orbitals, but with significant additional weight appearing on the surrounding bonding molecular oxygen orbitals. By explicitly comparing these results to those from the simpler single-band Hubbard model, our study provides strong support for the single-band framework for describing superconductivity in the cuprates.
\end{abstract}

\maketitle

\textit{Introduction} --- Cuprate superconductivity emerges in their quasi-two-dimensional (2D) CuO$_2$ planes after doping additional carriers into these layers. The undoped parent compounds are charge transfer insulators due to the large Coulomb repulsion $U_{dd}$ on the Cu 3$d$ orbitals, and, to a good approximation, a spin-$\tfrac{1}{2}$ hole is located on every Cu 3$d_{x^2-y^2}$ orbital. This situation is well described by a 2D square lattice Hubbard model or Heisenberg model in the large $U_{dd}$ limit. 

\begin{figure}[ht]
\centering
\includegraphics[width=0.8\linewidth]{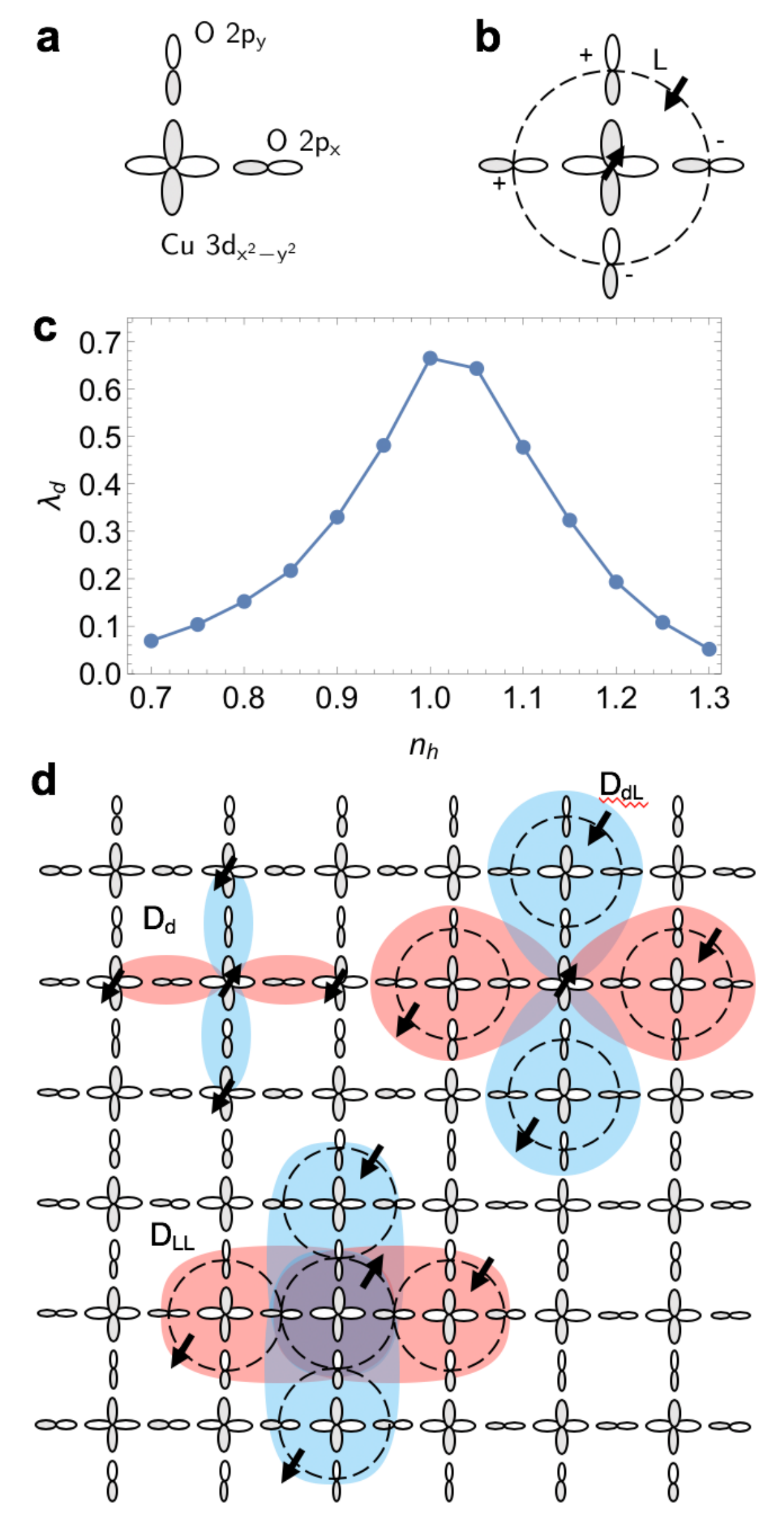}
\caption{{\bf a} The orbital basis of the three-band Hubbard model. {\bf b} Sketch of the bonding ligand ($L$) molecular orbital surrounding a central Cu-$d$ orbital. {\bf c} Leading BSE eigenvalue $\lambda_d$ vs $n_h$ for a $4\times4$ cluster at $\beta=16$ ~eV$^{-1}$. {\bf d} Sketches of some ways a pair can form with a $d$-wave symmetry.  Here, $D_d$ and $D_{dL}$ pair a Cu $3d$ hole with a hole on the neighboring Cu-$d$ and $L$ molecular orbital, respectively, while $D_{LL}$ pairs holes on neighboring O-$L$ orbitals.}
\label{Fig:PairStructure}
\end{figure}

Upon doping the additional holes or electrons primarily occupy the O or Cu orbitals, respectively. The minimal model capturing this asymmetry is the three-band Hubbard model, which explicitly accounts for the Cu $3d_{x^2-y^2}$, O $2p_{x}$, and $2p_y$ orbitals (Fig.~\ref{Fig:PairStructure}{\bf  a})~\cite{EmeryPRL1987}. Even at finite doping, the low energy sector of the three-band model can be mapped approximately onto an effective single-band Hubbard model~\cite{ZhangRice}. One expects this in the case of electron-doping since the additional carriers go directly onto the Cu sublattice, on which the holes of the undoped materials already reside. The case of hole-doping, however, is more subtle. Here, the additional carriers predominantly occupy the O sublattice due to the large $U_{dd}$ on the Cu orbital, and the appropriateness of a single-band model is less clear. In their seminal work, Zhang and Rice ~\cite{ZhangRice} argued that the doped hole effectively forms a spin-singlet state with a Cu hole, the ``Zhang-Rice singlet'' (ZRS, Fig.~\ref{Fig:PairStructure}{\bf b}), which then plays the same role as a fully occupied or empty site in an effective single-band model, again facilitating a single-band description.

The nature of the single-band 2D Hubbard model's pairing interaction has been extensively studied \cite{Maier4PRL, Maier07, Maier08, Kyung09, Gull14, Maier16}. Detailed calculations of its momentum and frequency structure using dynamic cluster approximation (DCA) quantum Monte Carlo (QMC) \cite{Maier4PRL} find that it is well described by a spin-fluctuation exchange interaction \cite{Maier07}. The single-band model, however, cannot provide any information on the orbital structure of the interaction. For example, in the hole doped case, the spins giving rise to the spin-fluctuation interaction are located on the Cu sublattice, while the paired holes are moving on the O $p_{x/y}$ sublattice. This situation can produce a different physical picture than if the interaction and the pairs both originate from the same orbital on the same lattice \cite{Lau11,Ebra14,Ebra16,Adolphs16,Jiang20}. And indeed, studies have observed two-particle behavior in a two sublattice system that is not observed in a one-lattice system \cite{Moeller12}. Moreover, an analysis of resonant inelastic x-ray scattering studies has found that a single-band model fails to describe the high-energy magnetic excitations near optimal doping \cite{Chen13}. Studying the effective interaction in a three-band model, and, in particular, determining its orbital structure is, therefore, critical. Such a study will also provide new insight into the nature of high-temperature superconductivity that is not available from the previous single-band studies. In this letter, we use a QMC-DCA method to explicitly calculate the orbital and spatial structure of the effective interaction in a realistic three-band CuO$_2$ model, and compare the results with those obtained from a single-band model.

\textit{Model and Methods} --- The three-band Hubbard model we study can be found in Refs.~[\onlinecite{Kung,supplement}]. We adopted a parameter set appropriate for the cuprates~\cite{Kung,Czyzyk,Johnston,Ohta} (in units of eV): the nearest neighbor Cu-O and O-O hopping integrals $t_{pd} = 1.13$, $t_{pp} = 0.49$, on-site interactions $U_{dd} = 8.5$, $U_{pp} = 0$, and charge-transfer energy $\Delta = \varepsilon_p -\varepsilon_d = 3.24$, unless otherwise stated. Since we use a hole language, half-filling is defined as hole density $n_h=1$ and $n_h>1~(<1)$ corresponds to hole (electron)-doping. A finite $U_{pp}$ only leads to small quantitative changes in the results (see Fig.~S4~\cite{supplement}) but worsens the sign problem significantly~\cite{Kung}. Therefore, we keep $U_{pp}=0$ for this study.

We study the single- and three-band Hubbard models using DCA with a continuous time QMC impurity solver~\cite{Jarrell,Maier1,Urs}. 
We determine the structure of the pairing interaction by solving the Bethe-Salpeter equation (BSE) in the particle-particle singlet channel to obtain its leading eigenvalues and (symmetrized) eigenvectors~\cite{supplement, Maier4PRL}. A transition to the superconducting state occurs when the leading eigenvalue $\lambda(T=T_c) = 1$, and the magnitude of $\lambda <1$ measures the strength of the normal state pairing correlations. The spatial, frequency, and orbital dependence of the corresponding eigenvector, which is the normal state analog of the superconducting gap, reflects the structure of the pairing interaction~\cite{Maier16,Maier4PRL}.

\textit{Results} --- Figure~\ref{Fig:PairStructure}{\bf c} shows the leading eigenvalue of the BSE for the three-band model as a function of hole concentration $n_h$ obtained on a $4\times 4$ cluster with $\beta=1/k_\mathrm{B}T=16$~eV$^{-1}$. We find that it always corresponds to a $d$-wave superconducting state~\cite{KirtleyReview} and is larger for hole-doping ($n_h>1$) compared to electron-doping ($n_h<1$). The latter observation suggests a particle-hole asymmetry in $T_\mathrm{c}$ consistent with experiments and prior studies of the single and two-band Hubbard models~\cite{Maier3,Macridin}. (Although $\lambda_d$ is largest at half-filling, we expect that it asymptotically approaches one as the temperature decreases but never actually cross one due to the opening of a Mott gap. We observe such behavior in explicit calculations on smaller three-band clusters, see Fig. S1~\cite{supplement}.)

We now analyze the spatial and orbital structure of the leading eigenvector $\phi_{\alpha\beta}({\bf k})$ ($\alpha$ and $\beta$ denote orbitals), 
by Fourier transforming $\phi_{\alpha\beta}({\bf k})$ to real space to obtain $\phi_{{\bf r}_\beta}({\bf r}_{\alpha})$, where ${\bf r}_\beta$ denotes the position of the orbital taken as the reference site. We employed a $6\times6$ cluster to allow for long-ranged pairing correlations at $T=0.1$~eV. While this relatively high temperature is needed to mitigate the Fermion sign problem, we have found that the leading eigenvector changes very slowly as the system cools (see Fig. S2)~\cite{supplement}. 
From here on, we focus on results obtained at optimal (15\%) hole- or electron-doping. We have obtained similar results for different cluster sizes and for finite $U_{pp}$ (see Figs. S3 and S4)~\cite{supplement}, indicating that our conclusions are robust across much of the model phase space. 

\begin{figure*}[ht]
\centering
\includegraphics[width=0.8\linewidth]{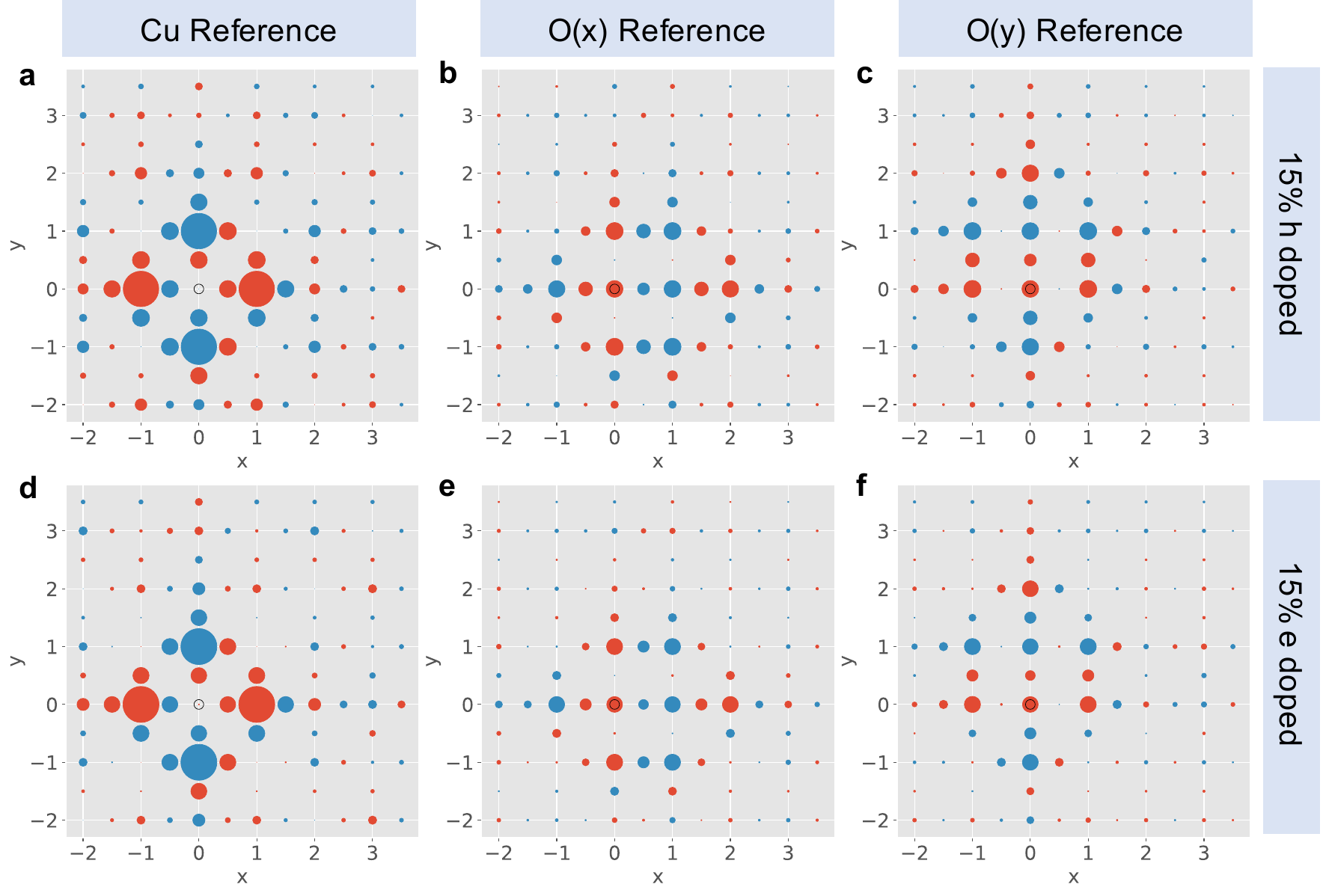}
\caption{The real space components of the leading  particle-particle BSE  (symmetrized) eigenvector for the three-band model at optimal doping and $\beta=10$ ~eV$^{-1}$ on a $6\times6$ cluster. Each column describes the pairing between a Cu $d$ (or O $p_x$, $p_y$) reference site and all other orbitals as a function of distance. All panels set the Cu-$d$ orbital at the origin, as labelled.}
\label{Fig:DCAPairStructure}
\end{figure*}

\begin{figure*}[ht]
\centering
\includegraphics[width=0.8\linewidth]{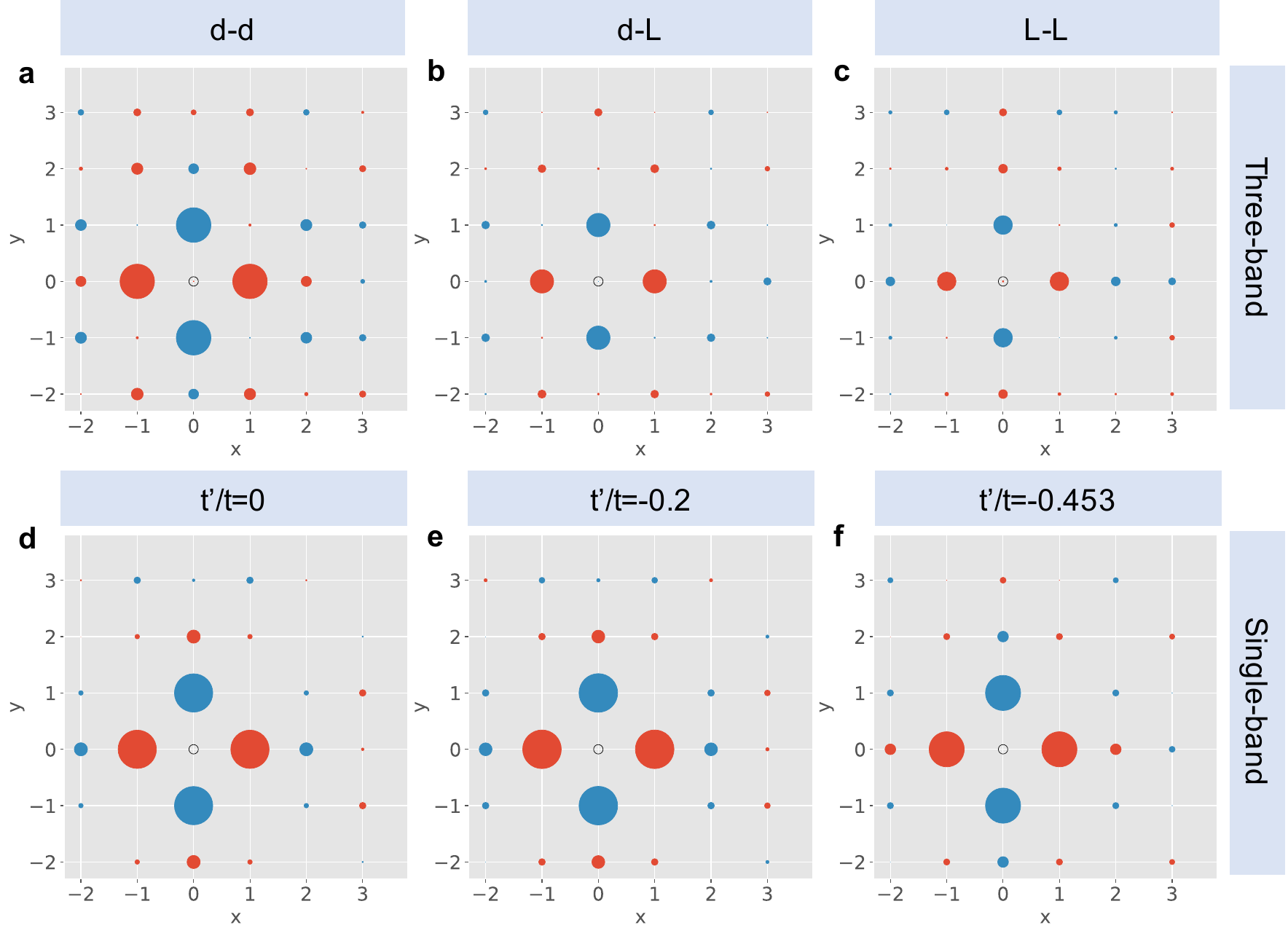}
\caption{A comparison of the orbital structure of the pairs in the three-band and single-band models at 15\% hole-doping. Each panel plots the real space components of the leading  particle-particle  BSE  (symmetrized) eigenvector. The first row shows $d$-$d$ $d$-$L$ and $L$-$L$ pairing for the three-band model at $\beta=10$ ~eV$^{-1}$. The second row shows the pair structure for the single-band model ($U=6t$, $\beta =5t^{-1}$) at $t^\prime/t=0$,$-0.2$,$-0.453$.}
\label{Fig:tbvssb}
\end{figure*}

In the single-band Hubbard model, the pairs are largely comprised of carriers on nearest neighbor sites in a $d$-wave state, i.e. with a positive (negative) phase along the $x$- ($y$)-directions. The internal structure of the pairs in the three-band model seems more complicated~\cite{Moreo}. The real-space structure of $\phi_{{\bf r}_\beta}({\bf r}_\alpha)$ shown in Figs.\ref{Fig:DCAPairStructure} {\bf a}-{\bf c} and Figs.\ref{Fig:DCAPairStructure} {\bf d}-{\bf f} for the hole- and electron-doped cases, respectively, display an extended and rich orbital structure. Here, the size and color indicate the strength and phase of $\phi_{{\bf r}_\beta}({\bf r}_\alpha)$, respectively, on each site after adopting the central Cu $3d_{x^2-y^2}$ or O $2p_{x,y}$ orbital as a reference site at ${\bf r}_\beta$. The form factors $\phi_{{\bf r}_\beta}({\bf r}_\alpha)$ are similar for both electron and hole doping, decaying over a length scale of $\sim 2$--$3$ lattice constants. Moreover, while the $d$-wave pairing between nearest Cu sites dominates, there is also a significant contribution from $d$-$p$ pairing, with a comparable amplitude for up to the third-nearest neighbors. The pairing between the individual O $2p_{x,y}$ orbitals is much weaker in comparison. 

We now transform the leading eigenfunction from the O-$p_x$ and $p_y$ basis to the bonding $L$ and anti-bonding $L^\prime$ combinations (Fig.~\ref{Fig:PairStructure}d). These combinations, formed from the four O orbitals surrounding a Cu cation, are the relevant states for the ZRS, in which the doped holes are argued to reside in. The bonding $L$ state strongly hybridizes with the central Cu 3$d_{x^2-y^2}$ orbital (Fig.~\ref{Fig:PairStructure}{\bf  b}), while the anti-bonding $L^\prime$ state does not. The resulting antiferromagnetic exchange interaction between the Cu and $L$ holes is then argued to bind them into the Zhang-Rice spin-singlet state, which provides the basis for the mapping onto a single-band model.

The orbital structure of the leading eigenfunction simplifies considerably after one transforms to the bonding $L$ and anti-bonding $L^\prime$ combinations. Fig.~\ref{Fig:tbvssb} plots the pairing amplitudes for a hole on Cu paired with another hole on a neighboring Cu ($d$-$d$, Fig.~\ref{Fig:tbvssb}{\bf a}) or bonding molecular orbital ($d$-$L$, Fig.~\ref{Fig:tbvssb}{\bf b}). Both components exhibit a clear $d_{x^2-y^2}$ symmetry; however, both channels also have indications of a higher momentum harmonic $[\cos(2k_xa)-\cos(2k_ya)]$. Interestingly, the contribution from holes occupying neighboring bonding molecular orbitals exhibits similar behavior ($L$-$L$, Fig.~\ref{Fig:tbvssb}{\bf c}). The $L^\prime$-related pairing contributes very little as will be discussed in Fig.~\ref{Fig:weightvsdensity} and in the supplement (see Fig. S5)~\cite{supplement}. 

Figs.~\ref{Fig:tbvssb}{\bf a}-{\bf c} establishes that the pairing between the different orbital components of the ZRS all possess the requisite $d_{x^2-y^2}$ symmetry. This indicates that the ZRS picture is valid for describing pairing correlations in the three-band Hubbard model of the cuprates. To confirm this, we also computed the real-space structure of the leading particle-particle BSE eigenfunction in the single-band Hubbard model. Here, we considered cases with next-nearest-neighbor hopping $t^\prime/t = 0$ (Fig.~\ref{Fig:tbvssb}{\bf d}) and $-0.2$ (Fig.~\ref{Fig:tbvssb}{\bf e}), which are commonly used in the literature, as well as $-0.453$ (Fig.~\ref{Fig:tbvssb}{\bf f}), which we obtained by downfolding our three-band model parameters onto the single-band model~\cite{Johnston2010,Eskes1991}. The single-band model reproduces the short-range pairing structure of the three-orbital model, regardless of the value of $t^\prime$; however, the second and third neighbor pairing is only captured correctly for $t^\prime/t=-0.453$. These results provide remarkable support for the validity of the ZRS construction but also indicate that single-band models may not capture the correct longer-ranged correlations without a suitable choice of $t^\prime$. The latter conclusion further underscores the crucial role of $t'$ for determining the superconducting properties of the single-band model~\cite{Maier16, Qin2019, Jiang2019}.

\begin{figure}[ht]
\centering
\includegraphics[width=0.7\linewidth]{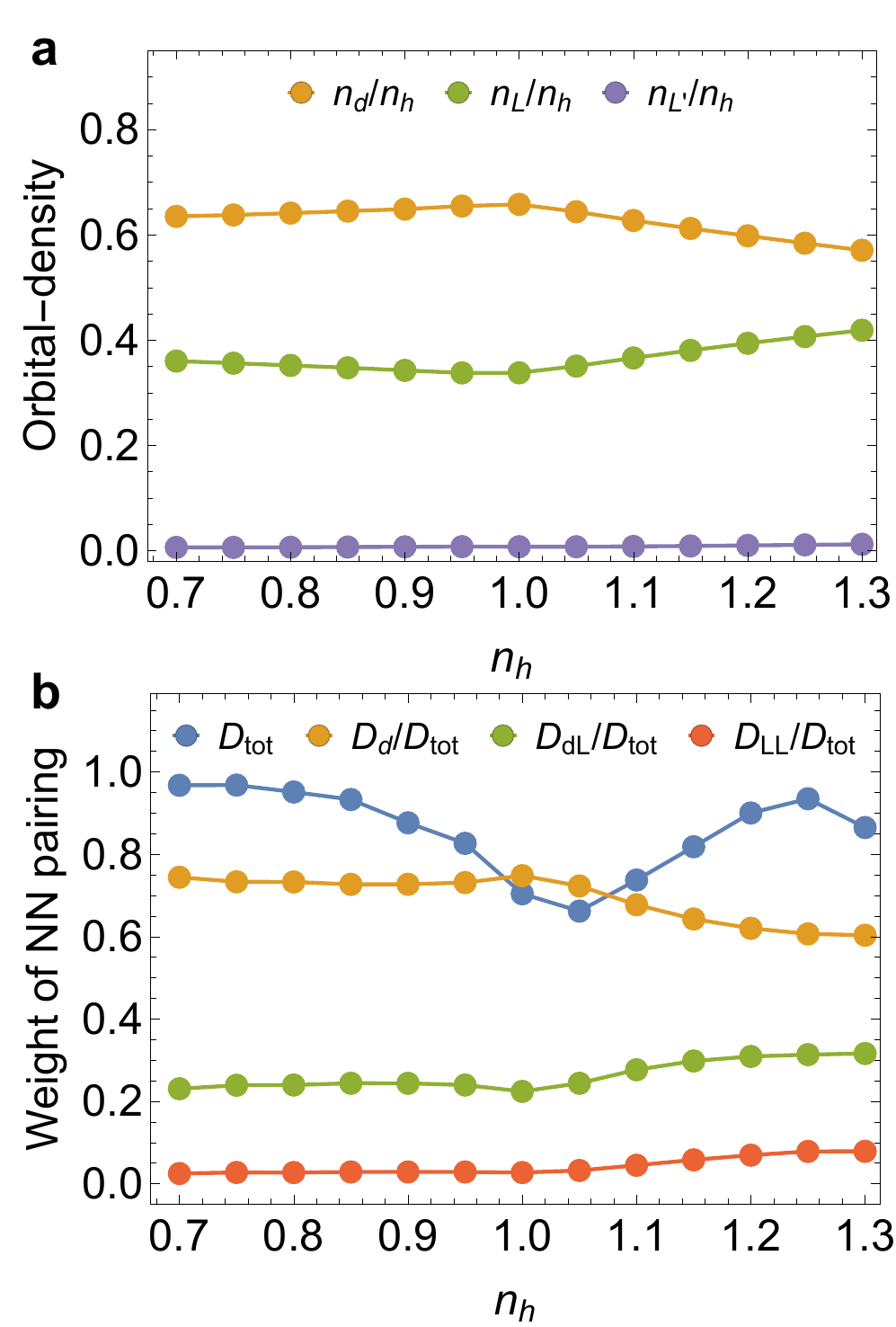}
\caption{{\bf a} Ratios of the orbital hole densities to the total density $n_h$. {\bf b} Weights of different orbital compositions of the nearest-neighbor pairs, as defined in Fig.\ref{Fig:PairStructure}, and their total weight. All results were obtained on a $4\times4$ cluster at $\beta=16$~eV$^{-1}$. 
}
\label{Fig:weightvsdensity}
\end{figure}

Figure~\ref{Fig:tbvssb} shows that the structure of the leading eigenfunction $\phi_{\alpha\beta}$ is closely linked to the orbital structure of the ZRS. Fig.~\ref{Fig:weightvsdensity} examines how this internal structure evolves with doping by plotting the orbital-dependent hole density (panel {\bf a}) and the orbital composition of the eigenfunction $\phi_{\alpha\beta}$ (panel {\bf b}) on a $4\times 4$ cluster (adequate to capture the essential pairing structure) at a lower temperature.  Fig.~\ref{Fig:weightvsdensity}{\bf a} shows that the single hole per unit cell in the undoped case has approximately 65\% Cu-$d$ character, while 35\% of the hole is located in the bonding O-$L$ molecular orbital. With electron doping, there is a small decrease of $n_d/n_h$ indicating that the holes are removed mainly from the Cu-$d$ orbital. In contrast, with hole doping, there is a significant redistribution of the hole density from the $d$- to the $L$-orbital, showing that doped holes mainly occupy the O-$L$ molecular orbital. The hole density on the anti-bonding O-$L^\prime$ orbital is negligible. 

Figure ~\ref{Fig:weightvsdensity}{\bf b} shows that the total weight of the nearest-neighbor pairing increases from $\sim 70$\% in the undoped case to almost 100\% with either hole or electron doping. Since the BSE eigenvector reflects the momentum structure of the pairing interaction, this dependence can be understood from an interaction that becomes more peaked in momentum space as $n_h=1$ is approached. This behavior leads to a more delocalized structure of $\phi_{{\bf r}_\beta}({\bf r}_\alpha)$ and, therefore, a reduction of the relative weight of the nearest-neighbor contribution. The partial contributions to the nearest-neighbor pairing weight, $D_{d}$ and $D_{dL}$, have a doping dependence very similar to the corresponding orbital densities $n_d$ and $n_L$ in panel {\bf a}, closely linking the orbital structure of the pairing to the orbital makeup of the ZRS. The weight of the $L^\prime$ contributions remains negligible over the full doping range~\cite{supplement}. 

\textit{Conclusions} --- We have determined the orbital structure of the effective pairing interaction in a three-band CuO$_2$ Hubbard model and shown that it simplifies considerably when viewed in terms of a basis consisting of a central Cu-$d$ orbital and a bonding $L$ combination of the four surrounding O-$p$ orbitals. These states underlie the ZRS singlet construction that enables the reduction of the three-band to an effective single-band model. By explicitly comparing the three-band with single-band results, we show that the effective interaction is correctly described in the single-band model. In summary, these results strongly support the conclusion that a single-band Hubbard model provides an adequate framework to understand high-$T_c$ superconductivity in the cuprates.


\begin{acknowledgments}
Acknowledgements --- The authors would like to thank L. Chioncel, P. Dee, A. Georges, K. Haule, E. Huang, S. Karakuzu, G. Kotliar, H. Terletska, and D.~J. Scalapino for useful comments. This work was supported by the Scientific Discovery through Advanced Computing (SciDAC) program funded by the U.S. Department of Energy, Office of Science, Advanced Scientific Computing Research and Basic Energy Sciences, Division of Materials Sciences and Engineering. This research  used resources of the Oak Ridge Leadership Computing Facility, which is a DOE Office of Science User Facility supported under Contract DE-AC05-00OR22725. The DCA++ code used for this project can be obtained at https://github.com/CompFUSE/DCA.
\end{acknowledgments}


\end{document}


\title{Orbital structure of the effective pairing interaction in the high-temperature superconducting cuprates -- Supplementary Material}

\author{Peizhi Mai$^1$, Giovanni Balduzzi$^2$, Steven Johnston$^{3,4}$, Thomas A. Maier$^{1,5,*}$}
\affiliation{$^1$Center for Nanophase Materials Sciences, Oak Ridge National Laboratory, Oak Ridge, TN 37831-6164, USA}
\affiliation{$^2$Institute for Theoretical Physics, ETH Z{\"u}rich, 8093 Z{\"u}rich, Switzerland}
\affiliation{$^3$Department of Physics and Astronomy, University of Tennessee, Knoxville, Tennessee 37996-1200, USA}
\affiliation{$^4$Joint Institute for Advanced Materials at The University of Tennessee, Knoxville, TN 37996, USA}
\affiliation{$^5$Computational Sciences and Engineering Division, Oak Ridge National Laboratory, Oak Ridge, TN, 37831-6494, USA}
\date{\today}


\flushbottom
\maketitle
\section{The single- and three-band Hubbard Models}
The Hamiltonian of the three-band Hubbard model is  
\begin{equation}
\begin{split}
H&=(\varepsilon_d-\mu)\sum_{i,\sigma}n_{i,\sigma}^d+ (\varepsilon_p-\mu)\sum_{j,\sigma} n^{p_\alpha}_{j,\sigma} 
+ \sum_{\langle  i,j\rangle,\sigma}t_{ij}(d^\dagger_{i\sigma}p^\pdag_{\alpha j\sigma}+p^\dagger_{\alpha  j\sigma}d^\pdag_{i\sigma})\\
&+\sum_{\langle j,j^\prime \rangle,\sigma} t_{jj^\prime}(p^\dagger_{\alpha j\sigma}p^\pdag_{\alpha^\prime j^\prime\sigma}+p^\dagger_{\alpha^\prime j^\prime\sigma}p^\pdag_{\alpha j\sigma}) + U_{dd} \sum_i n^d_{i\uparrow}n^d_{i\downarrow} + U_{pp} \sum_{j} n^{p_\alpha}_{j\uparrow}n^p_{j\downarrow}. 
\label{threeband}
\end{split} 
\end{equation}
Here, $d^\dagger_{i,\sigma}$ ($d^\pdag_{i,\sigma}$) creates (annihilates) a spin $\sigma$ ($=\uparrow,\downarrow$) hole in the copper $d_{x^2-y^2}$ orbital at site $i$;  $p^\dagger_{\alpha j\sigma}$ ($p^\pdag_{\alpha j\sigma}$) creates (annihilates) a spin $\sigma$ hole in the oxygen $p_\alpha$ ($\alpha = x,y$) orbital at site $j$; for nearest neighbor, $j=i\pm \hat{x}/2 \hspace{0.05cm} (\text{or} \hspace{0.1cm} \hat{y}/2)$; $n^d_{i\sigma}=d^\dagger_{i\sigma}d^\pdag_{i\sigma}$ and 
$n^{p_\alpha}_{j\sigma}=p^\dagger_{\alpha j\sigma}p^\pdag_{\alpha j\sigma}$ are the number operators; $\epsilon_d$ and $\epsilon_p$ are the onsite energies of the Cu and O orbitals, respectively; $\mu$ is the chemical potential; $t_{i,j}$ is the nearest neighbor Cu-O hopping integral; $t_{j,j^\prime}$ is the nearest neighbor O-O hopping integral; and $U_{dd}$ and $U_{pp}$ are the on-site Hubbard repulsions on the Cu and O orbitals, respectively. The hopping integrals are parameterized~\cite{Kung} as $t_{ij} = P_{ij}t_{pd}$ and $t_{jj^\prime} =  Q_{jj^\prime} t_{pp}$, where $P_{ij}=1$ for $j=i+\hat{y}/2$ or $j=i-\hat{x}/2$, $P_{ij}=-1$ for $j=i-\hat{y}/2$ or $j=i+\hat{x}/2$  and $Q_{jj^\prime}=1$ for $j'=j-\hat{x}/2+\hat{y}/2$ or $j'=j+\hat{x}/2-\hat{y}/2$, $Q_{jj^\prime}=-1$ for $j'=j+\hat{x}/2+\hat{y}/2$ or $j'=j-\hat{x}/2-\hat{y}/2$. Throughout, we adopted (in units of eV): $t_{pd} = 1.13$, $t_{pp} = 0.49$, $U_{dd} = 8.5$, $U_{pp} = 0$, and $\Delta = \varepsilon_p -\varepsilon_d = 3.24$~\cite{Kung,Czyzyk,Johnston,Ohta}, unless otherwise stated. Since we use a hole language, half-filling is defined as hole density $n_h=1$ and $n_h>1$ corresponds to hole-doping and $n_h<1$ corresponds to electron-doping. A finite $U_{pp}$ only leads to small quantitative changes in the pair structure (see Fig.~S4), but worsens the sign problem significantly \cite{Kung}. Therefore, we keep $U_{pp}=0$ for this study except for the results presented in Fig.~S4.

The downfolded single-band Hubbard model is 
\begin{equation}\label{singleband}
    H = -\mu\sum_{i \sigma} n_{i  \sigma} - \sum_{\langle  i,j\rangle \sigma} 
    t^\pdag_{ij}c^\dagger_{i \sigma}c^\pdag_{j \sigma} + 
    U \sum_{i}\left(n_{i \uparrow}-\frac{1}{2}\right)\left(n_{i \downarrow}-\frac{1}{2}\right),
\end{equation}
where $c^\dagger_{i \sigma}$ ($c^\pdag_{i \sigma}$) creates (annihilates) an electron with spin $\sigma$ at site $i$, $t_{i,j} = t$ and $t^\prime$ for nearest- and next-nearest-neighbor hoping, respectively, and zero otherwise. $U$ is the on-site Hubbard repulsion, and  $\mu$ is the chemical potential, which is adjusted to fix the electron filling. Throughout, we set $t = 1$, $U = 6t$, and vary $t^\prime$ as indicated in the text.

\section{Symmetrized eigenvectors of the Bethe-Salpeter equation}
To determine the structure of the effective pairing interaction, we solve the Bethe-Salpeter equation in the particle-particle singlet channel
\begin{equation} \label{eq:BSE}
    -\frac{T}{N_c} \sum_{K,\alpha_1,\alpha_2} \Gamma^{c, pp}_{\alpha,\beta,\alpha_1,\alpha_2}(K,K^\prime)\bar{\chi}_{\alpha_1,\alpha_2,\alpha_3,\alpha_4}(K^\prime)  \phi^{R,\nu}_{\alpha_3\alpha_4}(K^\prime) = \lambda_\nu \phi^{R,\nu}_{\alpha\beta}(K)\,.
\end{equation}
Here, $K=({\bf K},i\omega_n)$, and $\bar{\chi}_{\alpha_1,\alpha_2,\alpha_3,\alpha_4}({\bf K}, \omega_n) = (N_c/N) \sum_{{\bf k}^\prime} [G_{\alpha_1\alpha_3}({\bf K+k'},\omega_n)G_{\alpha_2\alpha_4}(-{\bf K}-{\bf k}^\prime, -\omega_n)]$ is the coarse-gained bare particle-particle propagator. The irreducible particle-particle vertex $\Gamma^{c, pp}$ is extracted from the two-particle cluster Green's function $G^{2,c}_{\alpha_1,\alpha_2,\alpha_3,\alpha_4}(K,K')$ with zero center of mass momentum and frequency by inverting the cluster Bethe-Salpeter equation
\begin{equation}
\begin{split}
G^{2,c}_{\alpha_1,\alpha_2,\alpha_3,\alpha_4}(K,K^\prime)&=\bar{G}_{\alpha_1,\alpha_3}(K)\bar{G}_{\alpha_2\alpha_4}(-K)\delta_{K,K^\prime}
\\&\
+\frac{T}{N_c}\sum_{K^{\prime\prime},\alpha^\prime_1\dots\alpha^\prime_4}\bar{G}_{\alpha_1,\alpha^\prime_1}(K)\bar{G}_{\alpha_2,\alpha^\prime_2}(-K)\Gamma^{c,pp}_{\alpha^\prime_1,\alpha^\prime_2,\alpha^\prime_3,\alpha^\prime_4}(K,K^{\prime\prime})G^{2,c}_{\alpha^\prime_3,\alpha^\prime_4,\alpha_3,\alpha_4}(K^{\prime\prime},K^\prime)\,.
\label{BSE1}
\end{split}
\end{equation}
To remove the ambiguity between left and right eigenvectors of the eigenvalue equation (\ref{eq:BSE}), we symmetrize the pairing kernel entering Eq.~(\ref{eq:BSE}). Using matrix notation in $(K, \alpha, \beta)$, we first diagonalize the bare particle-particle propagator, $\bar{\chi}^D = U^{-1}\bar{\chi}U$, where $\chi^D$ is a diagonal matrix, to introduce the symmetrized BSE
\begin{equation}\label{sBSE}
-\frac{T}{N_c} U \sqrt{\chi^D}U^{-1} \Gamma^{c,pp} U \sqrt{\chi^D}U^{-1} \phi^\nu = \lambda_\nu \phi^\nu\,.
\end{equation}
We use the eigenvectors of the symmetrized BSE, $\phi^\nu_{\alpha\beta}(K)$, for the analysis presented in the main text. They are related to the right eigenvectors of the BSE in Eq.~(\ref{eq:BSE}) by 
\begin{equation}
    \phi^\nu = U\sqrt{\chi^D}U^{-1}\phi^{R,\nu}\,.
\end{equation}

\section{The basis transformation to the molecular $L$, $L^\prime$ orbitals}
The construction of the Zhang-Rice singlet relies on a transformation from the oxygen $p_x$, $p_y$ orbital basis to bonding and anti-bonding molecular orbitals, denoted here 
as $L$ and $L^\prime$, respectively. The two basis are related by a unitary transformation \cite{ZhangRice,Avella2013,Maier4} defined in $k$-space  
\begin{equation}
L_{{\bf k}\sigma}=\frac{i}{\gamma_{{\bf k}}}~\left[\sin\left(\tfrac{k_xa}{2}\right)~p_{x{\bf k}\sigma}- \sin\left(\tfrac{k_ya}{2}\right)~p_{y{\bf k}\sigma}\right],\label{Lk}
\end{equation}
and
\begin{equation}
L^\prime_{{\bf k}\sigma}=\frac{-i}{\gamma_{{\bf k}}}~\left[\sin\left(\tfrac{k_ya}{2}\right)~p_{x{\bf k}\sigma}+ \sin\left(\tfrac{k_xa}{2}\right)~p_{y{\bf k}\sigma}\right],\label{Lbark}
\end{equation}
where $\gamma^2_{{\bf k}}=\sin^2(k_xa/2)+\sin^2(k_ya/2)$,  $p_{\alpha{\bf k}\sigma}=N^{-1/2}_c\sum_{j} p_{\alpha j \sigma}\exp(-i~{\bf k} \cdot {\bf R}_j)$, and we have set the lattice constant $a = 1$. In this basis, only the $L$ state hybridizes with the Cu-$d$ orbital, while the $L^\prime$ state only hybridizes with the $L$ state. The Fourier transform of the $L$ and $L^\prime$ orbitals to real-space is defined as $L_{i\sigma}=N^{-1/2} \sum_{\bf k}L_{{\bf k}\sigma} \exp(-i~{\bf k} \cdot {\bf R}_i)$, $L^\prime_{i^\prime\sigma}=N^{-1/2} \sum_{\bf k}L^\prime_{{\bf k}\sigma} \exp(-i~{\bf k} \cdot {\bf R}_{i^\prime})$ where $i^\prime=i+\hat{x}/2+\hat{y}/2$.

\section{Superconducting transition temperature in the three-band Hubbard model}
\begin{figure}[ht]
\centering
\includegraphics[width=0.5\linewidth]{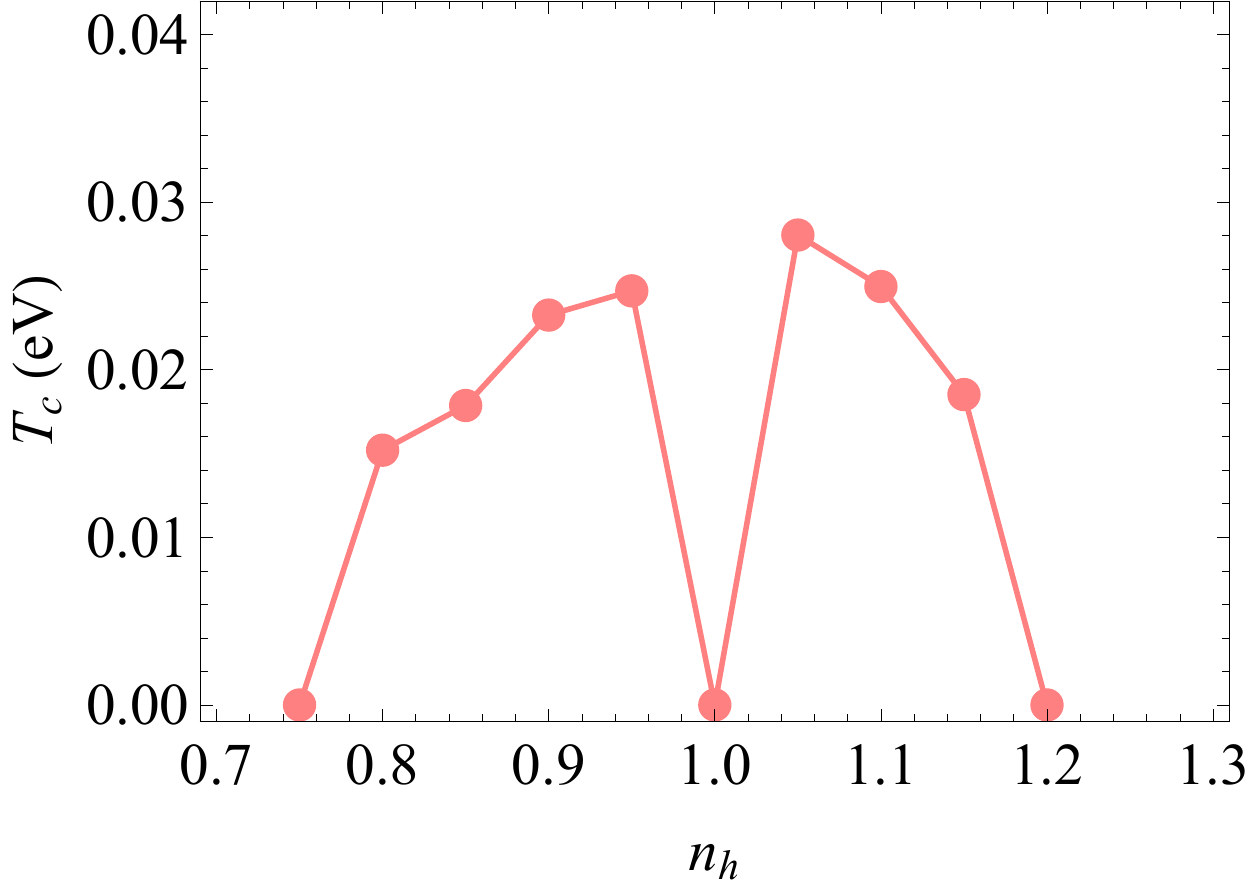}
\caption{{\bf Superconducting transition temperature as a function of filling for a $\mathbf{N_{Cu}=2\times2}$ DCA cluster}. The transition temperature $T_c$ is estimated by finding the temperature at which the leading BSE eigenvalue goes to $1$. The model parameters are (in units of eV) $\Delta = 3.24$, $t_{pd}=1.13$, $t_{pp}=0.49$, $U_{pp} = 0$, and $U_{dd}=8.5$. Our calculations find two superconducting domes on the hole-doped and electron-doped sides of the phase diagram, respectively, with a maximum $T_c$ that is higher for the hole-doped case, consistent  with experiments.}
\label{Fig:dcaTcvsn2by2Upp0SM}
\end{figure}

For the $6\times 6$ and $4\times 4$ DCA calculations presented in the main text, the QMC Fermion sign problem prevents calculations down to temperatures low enough to determine the superconducting transition temperature $T_c$ from the temperature where the leading eigenvalue of the Bethe-Salpeter equation crosses 1, i.e. $\lambda_d(T=T_c)=1$. This temperature can be reached on a $2\times 2$ cluster, however, and $T_c(n_h)$ can be determined as a function of hole density $n_h$ in that case. Fig.~\ref{Fig:dcaTcvsn2by2Upp0SM} shows the DCA results for $T_c(n_h)$ obtained in a $2\times 2$ cluster for the same model parameters as used in the main text. Similar to the electron-hole asymmetry found in Fig.~1 {\bf c} for the leading eigenvalue $\lambda_d(n_h)$ of the particle-particle Bethe-Salpeter equation, as well as the asymmetry found in experiments, the $T_c$ versus $n_h$ phase diagram exhibits a higher maximum $T_c$ on the hole doped side than on the electron-doped side. Moreover, these results are similar to previous DCA $2\times 2$ cluster calculations for a similar two-band model \cite{Macridin}, although the critical hole doping where $T_c$ vanishes is reduced compared to those earlier calculations. This difference may originate in the difference in model parameters, in particular the neglect of the direct oxygen-oxygen hopping $t_{pp}$ in the earlier two-band model calculations. 

\section{Dependence of the leading eigenfunction on temperature, cluster size and oxygen Coulomb repulsion}

\begin{figure}[h]
\centering
\includegraphics[width=\linewidth]{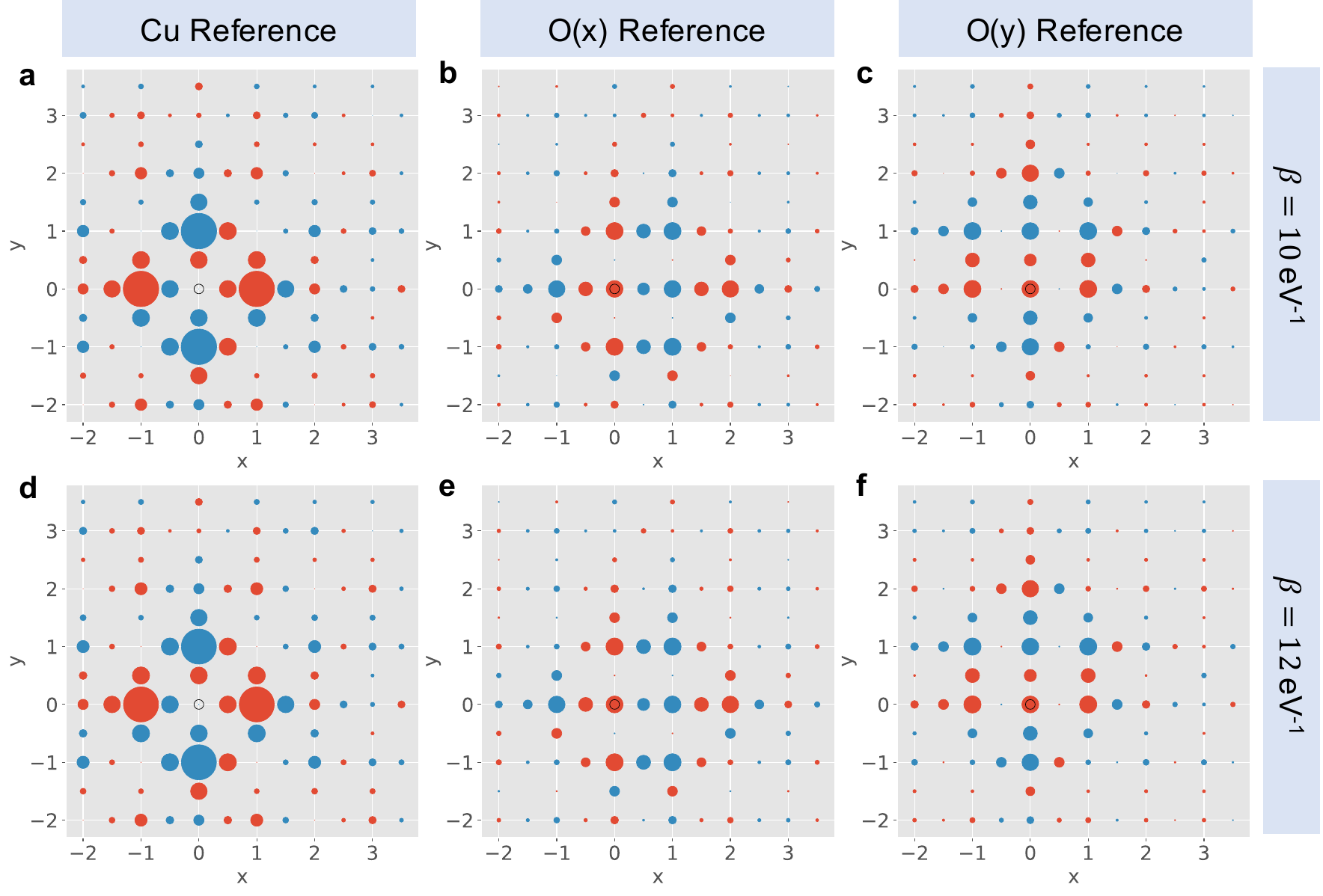}
\caption{{\bf Temperature dependence of the orbital structure of the pairs in the 
three-band model for the CuO\boldsymbol{$_2$} plane}. Each panel plots the real space components of the leading (symmetrized) eigenvector of the Bethe-Salpeter equation. The top and bottom rows show results obtained on a $N_{Cu}=6\times6$ cluster with a hole filling $n_h=1.15$ at an inverse temperature $\beta=10$ ~eV$^{-1}$ and $\beta=12$ ~eV$^{-1}$ respectively. The remaining model parameters are (in units of eV) $t_{pd}=1.13$, $t_{pp}=0.45$, $\Delta=3.24$, $U_{pp} = 0$, and $U_{dd}=8.5$. The left column describes the pairing between a Cu $d$ reference site and all other orbitals as a function of distance. The middle column describes pairings with respective to a p$_x$ orbital reference. The right column describes pairings with respective to a p$_y$ orbital orbital reference. All panels set the Cu $3d$ orbital at the origin, as labeled by the open ring. Only slight changes are observed in the pair structure between these two temperatures.}
\label{Fig:DCAeigenvector6by6varyT}
\end{figure}

While the leading eigenvalue $\lambda_d(T)$ shows a very strong increase with decreasing temperature, the temperature dependence of the corresponding eigenfunction is found to be rather weak. Fig.~\ref{Fig:DCAeigenvector6by6varyT} shows how the orbital and spatial structure of the leading eigenfunction $\phi_{{\bf r}_\beta}({\bf r}_\alpha)$ of the (symmetrized) Bethe-Salpeter equation changes with decreasing temperature between $\beta=1/T=10$~eV$^{-1}$ (top panels {\bf a}-{\bf c} from Fig.~2 in the main text) and $\beta=12$~eV$^{-1}$ (bottom panels {\bf d}-{\bf f}). We only observe small quantitative changes between these two temperatures. 

\begin{figure}[h]
\centering
\includegraphics[width=\linewidth]{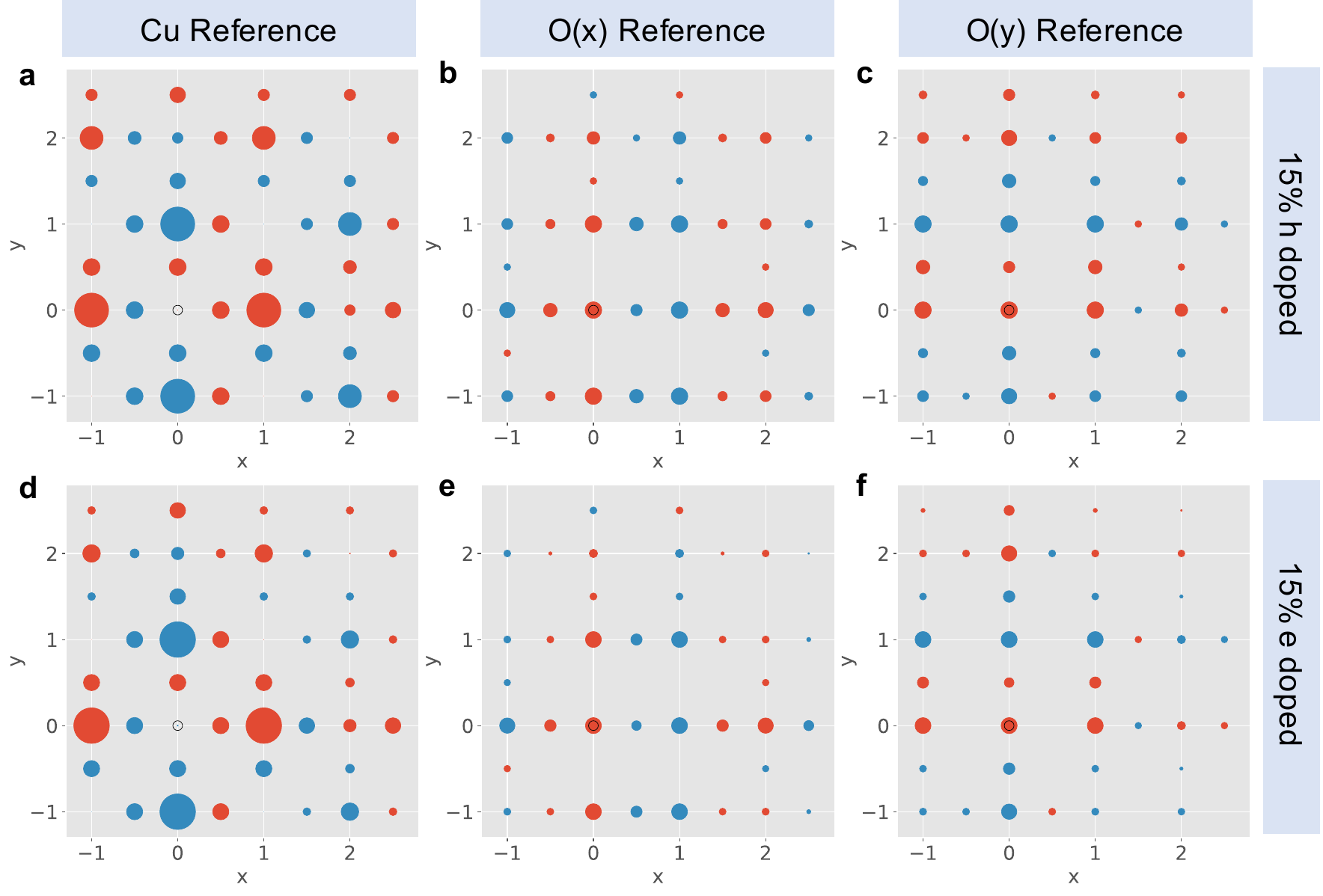}
\caption{\textbf{The orbital structure of the pairs in the 
three-orbital model for the CuO\boldsymbol{$_2$} plane.} Each panel plots the real space components of the leading (symmetrized) eigenvector of the Bethe-Salpeter equation.  The top and bottom rows show results for hole- ($n_h=1.15$) and electron-doping ($n_h=0.85$), respectively, obtained on $N_{Cu} = 4\times 4$ clusters and at an inverse temperature $\beta=16$ ~eV$^{-1}$. The remaining model parameters are (in units of eV) $t_{pd}=1.13$, $t_{pp}=0.45$, $\Delta=3.24$, $U_{pp} = 0$, and $U_{dd}=8.5$. The left column describes the pairing between a Cu $d$ reference site and all other orbitals as a function of distance. The middle column describes pairings with respect to a p$_x$ orbital reference. The right column describes pairings with respect to a p$_y$ orbital orbital reference. All panels set the Cu $3d$ orbital at the origin, as labeled by the open ring. Compared with Fig.~2 in the main text, the $4\times4$ cluster contains the same essential pair structure as the $6\times 6$ cluster and makes it possible to explore lower temperatures and stronger interactions.}
\label{Fig:DCAeigenvector4by4}
\end{figure}

The cluster size dependence of the leading eigenfunction is studied in Fig.~\ref{Fig:DCAeigenvector4by4}, which shows the results of an $N_{Cu} = 4\times 4$ cluster calculation for a 15\% hole doped and a 15\% electron doped system. These results should be compared with Fig.~\ref{Fig:DCAeigenvector6by6varyT} (or Fig.~2 in the main text), which displays the same calculation for a larger $N_{Cu}=6\times 6$ cluster. From this comparison, one sees that the $4\times 4$ cluster is large enough to contain the important components of the eigenfunction. Since the Fermion sign problem is much less severe in the $4\times 4$ cluster, it allows for calculations at lower temperatures or with an additional on-site Coulomb repulsion $U_{pp}$ on the oxygen orbitals. 

\begin{figure}[h]
\centering
\includegraphics[width=\linewidth]{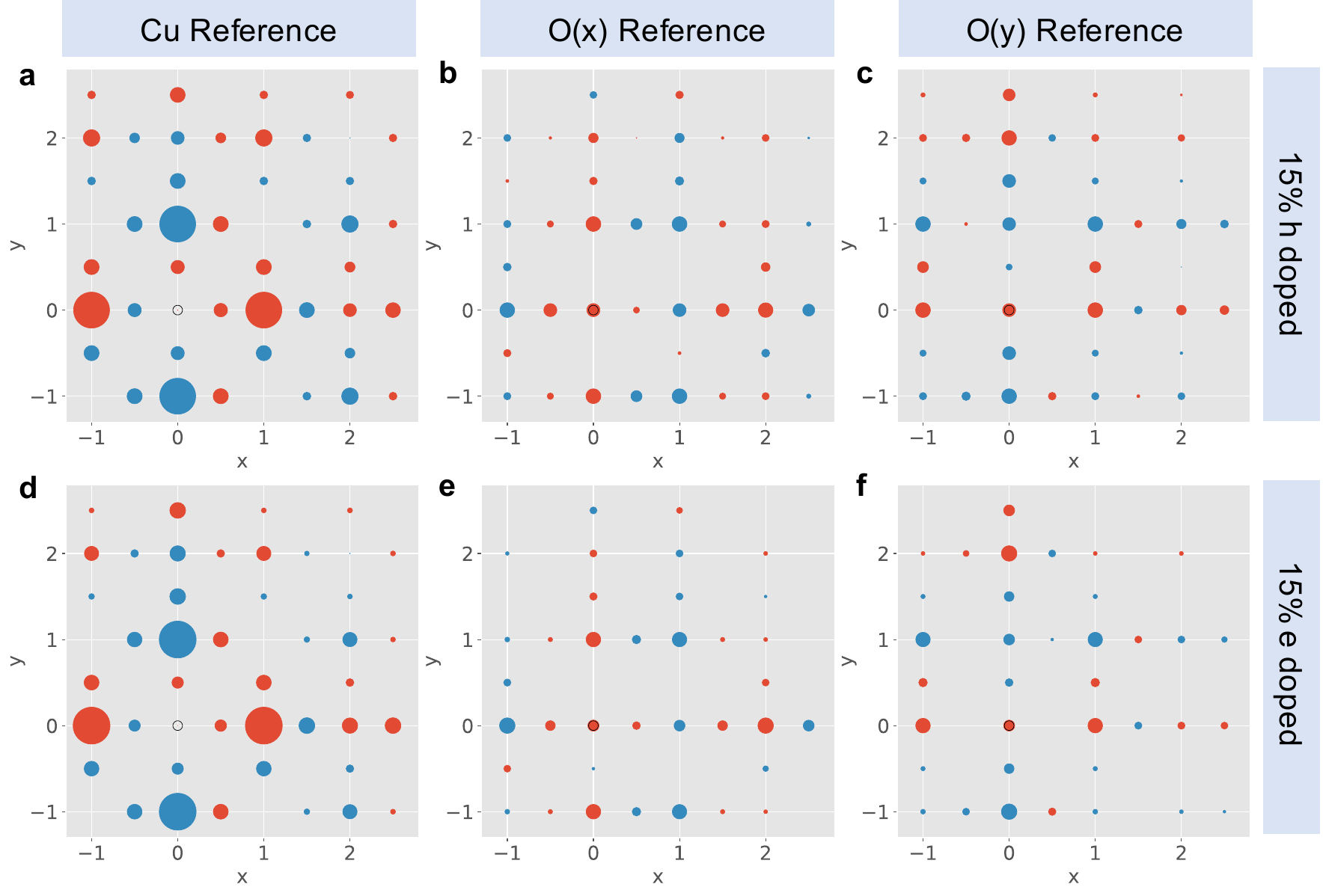}
\caption{\textbf{Effect of finite $\mathbf{U_{pp}}$ on the orbital structure of the pairs in the three-band model with finite oxygen Coulomb repulsion $\mathbf{U_{pp}}$.} Each panel plots the real space components of the leading (symmetrized) eigenvector of the Bethe-Salpeter equation. The first and second rows show results for hole- ($n_h=1.15$) and electron-doping ($n_h=0.85$), respectively, obtained on $N_{Cu} = 4\times 4$ clusters and at an inverse temperature of $\beta=10$ ~eV$^{-1}$ and finite $U_{pp}=4.1$. The remaining model parameters are (in units of eV) $t_{pd}=1.13$, $t_{pp}=0.45$, $\Delta=3.24$ and $U_{dd}=8.5$. Compared to Fig.~\ref{Fig:DCAeigenvector4by4}, the effect of a finite $U_{pp}$ is very weak, with only a very slight suppression of the components that involve the O-$p$ orbitals. }
\label{Fig:DCAeigenvector4by4Upp41}
\end{figure}

An additional $U_{pp} = 4.1$~eV term is considered in the data for the pair structure shown in Fig.~\ref{Fig:DCAeigenvector4by4Upp41}. Other model parameters are unchanged from those considered in Fig.~\ref{Fig:DCAeigenvector4by4}. Comparing these images with those in Fig.~\ref{Fig:DCAeigenvector4by4}, one sees that the structure of the eigenfunction remains almost unchanged by the additional $U_{pp}$. Only a very slight suppression of the components that involve the O-$p$ orbitals is observed. This justifies the neglect of $U_{pp}$ in most of our calculations, and provides evidence that our main conclusions reached from those calculations are general and not affected by $U_{pp}$. 

\section{Role of the anti-bonding molecular $\mathbf{L^\prime}$ orbital}

\begin{figure}[ht]
\centering
\includegraphics[width=\linewidth]{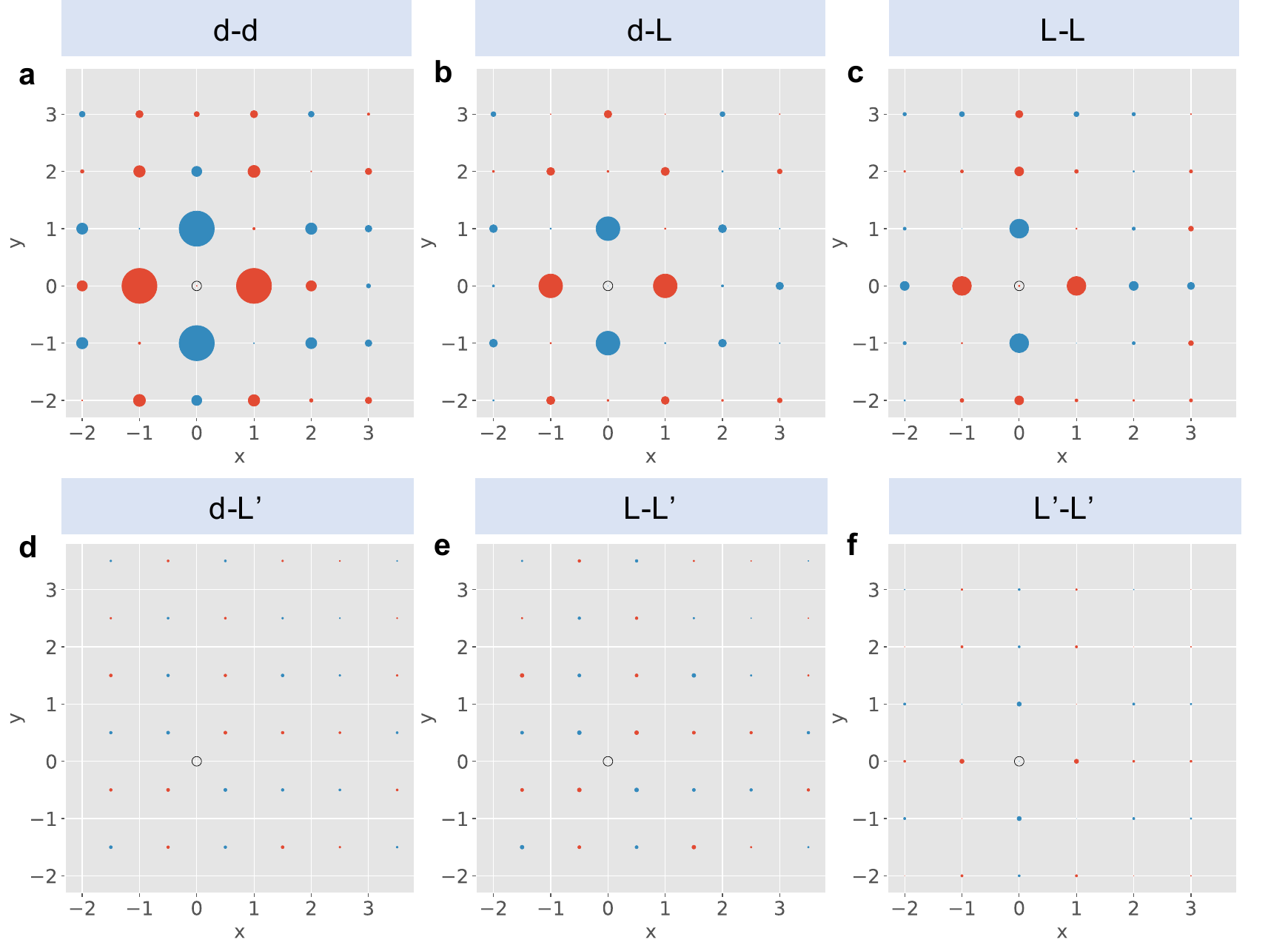}
\caption{\textbf{The $\mathbf{L^\prime}$-related components of the Cooper pairs in the three-band model for the CuO\boldsymbol{$_2$} plane.} $d$-$L^\prime$, $L$-$L^\prime$, $L^\prime$-$L^\prime$ pairing components are presented in panel {\bf d}, {\bf e}, {\bf f}, respectively, as compared with Panel {\bf a}-{\bf c} from Fig.3 in the main text. All results were obtained  on a $N_{Cu} = 6\times6$ cluster with a filling $n_h=1.15$ and at an inverse temperature $\beta=10$ ~eV$^{-1}$. The remaining model parameters are (in units of eV) $t_{pd}=1.13$, $t_{pp}=0.49$, $U_{pp} = 0$, and $U_{dd}=8.5$. The same scale is used for the size of the points in the top and bottom rows. The pairing with the $L^\prime$-orbital is negligible.}
\label{Fig:weightvsdelta}
\end{figure}

Finally, we show the components of the leading eigenfunction that involve the antibonding $L^\prime$ orbital in the bottom row of Fig.~\ref{Fig:weightvsdelta}, compared to the bonding $L$ components that were already shown in Fig.~3 of the main text. From the results for the orbital hole densities in Fig.~4 {\bf a} of the main text, it is clear that the $L^\prime$ molecular orbital remains almost completely unoccupied over the entire doping range considered, despite the finite hybridization between the $L$ and $L^\prime$ states. As a consequence, and as seen from the bottom row of Fig.~\ref{Fig:weightvsdelta}, the $L^\prime$ state is not involved in the pairing. This provides strong support for the Zhang-Rice picture, which only considers the $d$- and $L$-states in the mapping to an effective single-band model. 

